\documentclass[12pt]{amsart}
\usepackage{amscd}
\usepackage{graphicx}
\newtheorem{theorem}{Theorem}[section]

\newtheorem{defn}[theorem]{Definition}
\newtheorem{rem}[theorem]{Remark}

\title[MQTC Using Quantum Fourier Transforms]{Multiparty Quantum Telecommunication Using Quantum Fourier Transforms}
\author[Do Ngoc Diep]{Do Ngoc Diep${}^{1,2}$}
\address{${}^1$ Institute of Mathematics, Vietnam National Academy of Science and Technology, Hoang Quoc Viet road, Cau Giay district, 10307 Hanoi, Vietnam}
\email{dndiep@math.ac.vn}
\address{${}^2$ Thang Long University, Nghiem Xuan Yem, Hoang Mai district, Hanoi, Vietnam}
\begin{document}
\maketitle
\begin{abstract}  
Consider the problem: Alice wishes to send the same key to $n-1$ users (Bob, Carol,. . . , Nathan), while preventing eavesdropper Eve from acquiring information without being detected. The problem has no solution in the classical cryptography but in quantum telecommunication there are some codes to solve the problem.
In the paper \cite{zengetall}, Guo-Jyun Zeng, Kuan-Hung Chen, Zhe-Hua Chang, Yu-Shan Yang, and Yao-Hsin Chou from one side and Cabello in \cite{cabello} from other side, used Hadamard gates, Pauli gates in providing the quantum communication code for two-partity telecommunication with 3 persons and then generalized it to the case of arbitrary number of participants, indicating the position of measurements of participants. We remark that the Hadamard gate with precising the position of measurement is the same as Fourier transform for two qubits and hence use the general Fourier transform for $n$ entangled qubits, in place of Hadamard gates. The result is more natural for arbitrary $n$ qudits.
\end{abstract}

\section{Introduction}
The paper is devoted to the following problem of multiparty quantum telecommunication, see \cite{cabello} for more details:
Alice wishes to send a sequence of random classical bits (a “key”) to Bob, and at the same time  preventing that Eve acquires information without being detected.

This problem, as known has no solution by classical cryptography, but it can be solved in quantum telecommunication by using the quantum computing. There are some tools to solve the problems: Some peoples use the non-cloning principle, some ones use entanglement particles, some  others  combine quantum techniques with classical private amplification and compression techniques, or split out the information in several qubits to which eavesdropper Eve has only a sequential access.

In the works \cite{cabello} and \cite{zengetall} the problem was solved by a procedure that we will remind in the next sections 2. 
In those code, the Hadamard gates were used to makes measurement of entangled Bell states and/or general entangled GHZ-states some times with indication of position to make measurements and directions\cite{zengetall}. We remark that in the last the Hadamard gates \frame{H} can be replaced by a more general quantum Fourier transform \frame{$F_n$} and the Fourier coefficients can be considered as the results of measurement, which is called \textit{quantum Fourier measurements}.
In section 3 we show how to measure the quantum Fourier transform coefficients. and in section 4 we show how to use it to solve the general problem of multipartite quantum telecommunication. Our main results are Definition \ref{defn1} and Theorem \ref{thm1}. In section 5 we remind of security estimation and in Section 6 we illustrate the codes with the holographic softwares and finally in Section 7 we make some final conclusion is made.

\section{The  two-party quantum telecommunication}
The problem of two-party quantum teleportation between 3 persons can be formulated as follows. 
There three persons, Alice and Bob and Carol, who want to exchange their idea, Alice is some leader and Bob and Carol are participants.
In other word, Alice, Bob and Carol have the
secret key $S_A$, $S_B$ and $S_C$ respectively. And the final key is
$S_A \oplus S_B\oplus S_C$ and that each
participant can change the key by their idea,  see \cite{cabello}, \cite{zengetall} for more details.

In practice the problem appears in some specialized context, namely \cite{cabello}:

\textit{A. Hillery-Bužek-Berthiaume secret sharing using
GHZ states} 
 ``Alice wants to have a secret action taken on her behalf in a distant part. There she has two agents, Bob and Carol, who carry it out for her. Alice knows that one and only one of them is dishonest, but she does not know which one. She cannot simply send a secure message to both of them, because the dishonest one will try to sabotage the action, but she knows that if both carry it out together, the honest one will keep the dishonest one from doing any damage.''

\textit{B. Multiparty key distribution based on  Bell-state entanglement swapping and and Pauli actions} 
Supposed to consider the same situation but produce another protocol for secret sharing using
Bell states instead of GHZ states is proposed, but also some Pauli gates $\sigma_z$, $\sigma_x$. This was later develloped in the work \cite{zengetall}. 

\textit{C. Secret sharing using Bell-state and GHZ entanglement swapping} 
Supposed to consider the same situation but
produce a protocol such that once Bob  and Carol knows the results of
the public measurement, he/she can infer the first bit
of the result of Alice’s secret measurement and then use the public measurements from Alice they in cooperating together, can infer the second.

The four steps solution of the problem from can be summerized as follows, see \cite{cabello} for more details.
$$\CD 
\begin{array}{l}
\mbox{\bf Step i. Initialization:} \mbox{Alice uses 5 qubits: 1 and 2}\\ \mbox{(unmoving) in Bell state, and  3 (unmoving),  A and B }\\
\mbox{(moving for interchange) in GHZ state} \\
 \mbox{Carol has 2 qubits: 4(unmoving) and C(moving) in Bell state}\\
  \mbox{Bob has 2 qubits: 5(unmoving) and D (moving  ) in Bell state}\end{array}\\
@VV\begin{array}{l}
\mbox{Alice measures the Bell state in qubits 1 and 2, }\\
\mbox{Alice measures GHZ state in qubits 3, A and B, }\\
\mbox{Carol measures the Bell state in qubits 4 and C}\\
 \mbox{Bob measures the Bell state in qubits 5 and D}\\
 \end{array}V\\
 \begin{array}{l}
\mbox{After that the system is for example, namely in the initial state:}\\
|\psi_i\rangle = |000\rangle_{3AB} \otimes |00\rangle_{12} \otimes |00\rangle_{4C} \otimes |00\rangle_{5D} \\
\\
\mbox{\bf Step ii.  Bell-state measurements}
 \end{array} \\
@VV\begin{array}{l}\mbox{- Alice sends qubit A to Bob, qubit B to Carol}\\
 \mbox{- Bob performs a  Bell-state measurement}\\ \mbox{ on qubits 5 and D }\\
 \mbox{- Carol performs a  Bell-state measurement}\\ \mbox{ on qubits 4 and C }
 \end{array}
V \\
\begin{array}{l}
\mbox{After that it is easy to see that the system is in the state:} \\
|\psi_{ii}\rangle = |AP\rangle_{12}\otimes |BP\rangle_{5D} \otimes |CP\rangle_{4C} \\
\\
\mbox{\bf Step iii. Secret Bell-state measurements.}
\end{array}\\
@VV\begin{array}{l}\mbox{- Alice performs a \textit{secret} Bell-state measurement}\\ \mbox{ on qubits 2 and 3}\\
 \mbox{- Bob performs a \textit{secret} Bell-state measurement}\\ \mbox{ on qubits 5 and A }\\
 \mbox{- Carol performs a \textit{secret} Bell-state measurement}\\ \mbox{ on qubits 4 and B }\end{array}V \\
\begin{array}{l}
\mbox{ After that it is easy to see that the system is in the state:}\\
|\psi_{iii}\rangle= |AP\rangle_{1CD} \otimes |AS\rangle_{23} \otimes |BS\rangle_{5A}\otimes |CS\rangle_{4B}\\
\\
\mbox{\bf Step iv. Secret sharing}
\end{array}\\
@VV\begin{array}{l}
\mbox{- Bob (resp., Carol) sends qubit D (resp., C) out}\\
\mbox{ to Alice}\\ \mbox{ - Alice performs a complete GHZ-state}\\  \mbox{measurement on qubits 1, C, and Dand publishes}\\
\end{array}V \\
\begin{array}{l}
\mbox{After that it is easy to see that the system is in the state:}\\
|\psi_{iv}\rangle = |AP\rangle_{1CD} \otimes |AS\rangle_{23} \otimes |BS\rangle_{5A} \otimes |CS\rangle_{4B}\\
\end{array}
\endCD$$

This four steps are illustrated on Figure 1.

\begin{figure}[!ht]
			\centering 
				\includegraphics[scale=0.4]{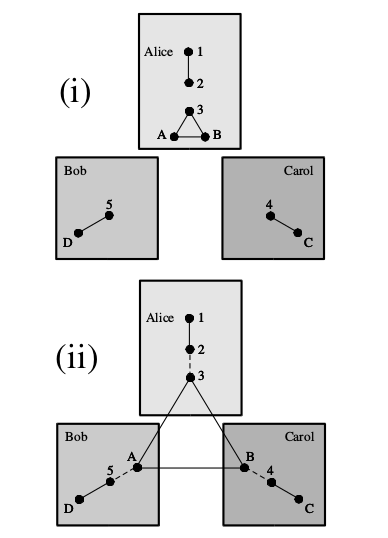}
                     \hspace{1cm}
                     \includegraphics[scale=0.4]{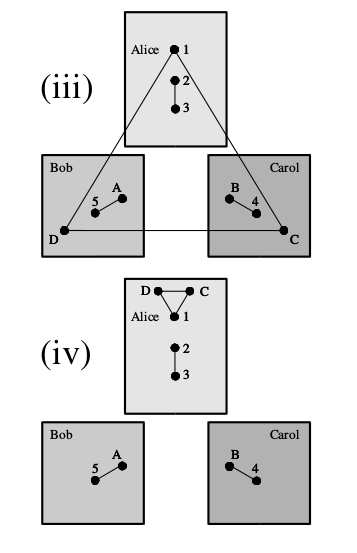}
                     \label{fig:1}
                     \caption{Steps 1-4: }
\end{figure}

\section{Quantum Fourier Transforms as Quantum Measurements}

The main tools are the entangled Bell state measurements and the entangled GHZ state measurements, which can be described as follows \cite{zengetall}.

\begin{figure}[!ht]
			\centering
			\label{fig:Bellmeasurement}
				\includegraphics[scale=0.4]{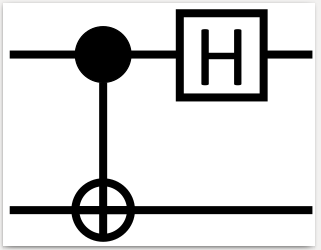}
			\hspace{1cm}
				\label{fig:GHZmeasurement}
				\includegraphics[scale=0.4]{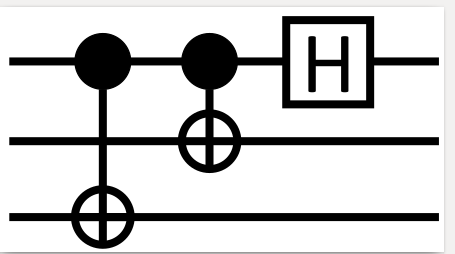}
                     \caption{The entangled measurement (a) The entangled Bell states measurement (b) The entangled GHZ states measurement.}
\end{figure}

We remark that the schemes for entangled Bell state measurments are the same as the Fourier transform \frame{$F_2$} on these two qubits and entangled GHZ state measurments are the same as the Fourier transform \frame{$F_3$} on three entangled qubits.
The next section is devoted to the general measurements of entangled 	GHZ states, using the quantum Fourier transforms  \frame{$F_n$} for and arbitrary number $n$ of partites.

For a state $|x\rangle$ we define the Fourier coefficients in a standard basis as the results of Fourier measurement.

\begin{defn}\label{defn1}
For any set of entangled qubit states $$|\mathbf x\rangle = \frac{1}{\sqrt{2^{n}}}\sum_{y\in F_2^{n}} e^{2\pi i\mathbf x.\mathbf y/2^{n}}|\mathbf y\rangle, $$ the Fourier coefficients are considered as the \textbf{results of Fourier measurement}.
\end{defn}

\begin{rem}
For a fixed orthonormal  basis consisting of $2^n$ vectors of $n$ entangled qubits in GHZ states \label{basis}
$$|0\dots 0\rangle_{ij\dots n}=\frac{1}{\sqrt{2}}\left(|0\rangle_i \otimes |0\rangle_j\otimes\dots\otimes |0\rangle_n + |1\rangle_i \otimes |1\rangle_j\otimes\dots \otimes|1\rangle_n\right)$$
$$|0\dots 1\rangle_{ij\dots n}=\frac{1}{\sqrt{2}}\left(|0\rangle_i \otimes |0\rangle_j\otimes\dots\otimes |0\rangle_n - |1\rangle_i \otimes |1\rangle_j\otimes\dots \otimes|1\rangle_n\right)$$
\dotfill
$$|11\dots 10\rangle_{ij\dots n}=\frac{1}{\sqrt{2}}\left(|1\rangle_i \otimes |1\rangle_j\otimes\dots\otimes |0\rangle_n + |0\rangle_i \otimes |1\rangle_j\otimes\dots \otimes|1\rangle_n\right)$$
$$|11\dots 11\rangle_{ij\dots n}=\frac{1}{\sqrt{2}}\left(|1\rangle_i \otimes |0\rangle_j\otimes\dots\otimes|0\rangle_n - |0\rangle_i \otimes |1\rangle_j\otimes\dots \otimes|1\rangle_n\right)$$
the Fourier transform measurements give the coefficients as the values of measurements.

\end{rem}

Let us first remind  the Quantum Fourier transform code picture, Figure 3 from Ekert lectures \cite{ekertetall}:

\begin{figure}[!ht]
			\centering
				\includegraphics[scale=0.4]{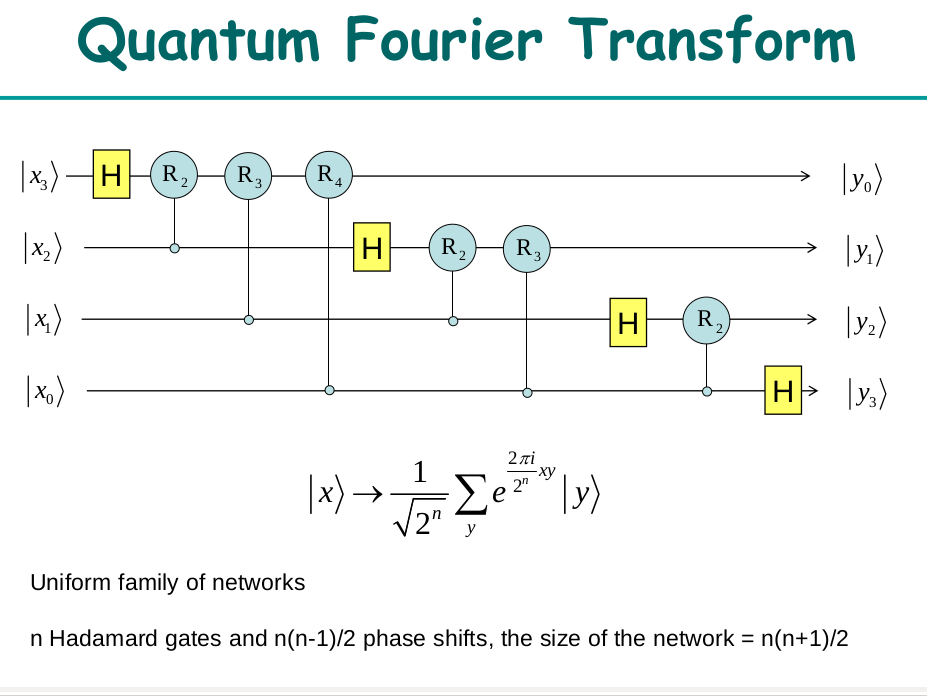}
                     \caption{Quantum Fourier transform measurement of multiparty entangled states}
\end{figure}

\section{The general case}
Consider the following problem: Alice wishes to convey the same key to N users (Bob, Carol,. . . , Nathan), while preventing Eve from acquiring information without being detected. This problem, called multiparty key distribution, is a special case of networked cryptographic
conferencing. Here I introduce a protocol for using GHZ states for
multiparty quantum key distribution that, as far as I know, has not been presented anywhere before. It can be considered as a generalization to many parties of the two-party protocol.

\begin{theorem} \label{thm1}
The above multiparty quantum telecommunication problem of secret sharing with quantum key distribution can be solved by a procedure with using the quantum Fourier transform measurements. 
\end{theorem}
\textit{Proof.} The theorem is proved by the 
 following 
 
 \textbf{Procedure}, the same as in the 3 persons case \cite{cabello}, the system state is changing as follows.
 $$\CD |\psi_i\rangle @>>> |\psi_{ii}\rangle @>>> |\psi_{iii}\rangle @>>> |\psi_{iv}\rangle \endCD$$ Let us consider in more details.
  
\textit{Step 1. Initialization of 3n qubits.} Alice has $n+2$ qubits: $1,2,3, A_1,\dots,  A_{n-1}$: qubits 1 and 2 are entangled in Bell state, qubits $3, A_1, \dots, A_{n-1}$ are entangled in GHZ state. $n-1$ persons: Bob, Carol, ...., Nathan, each has 2 entangled qubits $i+3, C_i,i=1\dots,n-1$ namely in null state. Alice produces a Bell state measurement on qubit 1 and 2 and a Fourier measurement \frame{$F_n$} on $n$ qubits $3, A_1,\dots, A_{n-1}$. Each of participants makes a Bell state Fourier measurement  \frame{$F_2$} of entangled $i+3,C_i, i=1\dots n-1$. 
At the end of this step 1, the system is in the state
$$|\psi_i\rangle = |0\dots 0\rangle_{3A_1\dots A_{n-1}} \otimes |00\rangle_{12} \otimes |00\rangle_{4C_1}\otimes\dots \otimes |00\rangle_{n+2,C_{n-1}} $$

\textit{Step 2. Entangled Bell-state measurements.}
Alice sends each qubit $A_i$  of her GHZ state out to
each $i^{th}$ participant of the other $n-1$ users. 
The system is in the state
$$|\psi_{ii}\rangle = |AP\rangle_{3A_1\dots A_{n-1}} \otimes |BP\rangle_{1C_1} \otimes|CP\rangle_{2C_2} \otimes\dots \otimes |NP\rangle_{{n+2},C_{n-1}} $$

\textit{Step 3. Secret Bell-state measurement.}
Next, Alice and each user  performs a Bell-state Fourier measurement  \frame{$F_2$} on the received qubit and one of their qubits.
After these measurements the state of the system
becomes
$$|\psi_{iii}\rangle = |AP\rangle_{3A_1\dots A_{n-1}} \otimes |AS\rangle_{2,3}\otimes |BS\rangle_{4,A_1} \otimes \dots \otimes |NS\rangle_{n+2,A_{n-1}},$$
where $|AP\rangle$ is $n$-qubit GHZ state of the standard basis \ref{basis}.

\textit{Step 4. Secret sharing.} 
The $n-1$ users sends a qubit (the one they have not used) to Alice, and she performs a Fourier measurement  \frame{$F_n$} to discriminate between the $2^n$ GHZ states, and publicly announces the result $|AP\rangle_{3C_1\dots C_{n-1}}$.
After these measurements the state of the system
becomes
$$|\psi_{iv}\rangle = |AP\rangle_{1C_1\dots C_{n-1}} \otimes |AS\rangle_{2,3}\otimes |BS\rangle_{4,A_1} \otimes \dots \otimes |NS\rangle_{n+2,A_{n-1}},$$
  The result AP, and the result of their own secret measurement allow each legitimate user to infer the first bit of Alice’s secret result $AS$. To find out the second bit of AS, all users (except Alice) must cooperate. The proof therefore is achieved. \hfill$\Box$

\begin{rem}
The same is true for qudits in place of qubits. In that situation we do use the phase Fourier coefficients in place of $\pm$.
\end{rem}

\section{Security}
It was shown \cite{cabello} that the protocools guarantees the security in the following sense: The secret that Alice admits has two qubits 2 and 3. by the public entangled Bell-state measurement  of qubit 1 and 2, 4 and $C_1$, 5 and $C_2$ etc. and the entangled GHZ state of 1 and $C_1$ and $C_{n-1}$, every participant knows the first bit of the secret of Alice.

In order to find out the second bit of Alice's secret all $n-1$ participand do cooperate together and following the public result of GHZ state measurement of qubits $3, A_1,\dots,A_{n-1}$ every body knows also the second bit of the qubit 3 of Alice. The Alice's secret is therefore discovered by each participant.  

A eavesdropper Eve can not do  change the situation: Eve need to have the same as each participant - any attempt ot find out one of the secret result of participant will change the Alice's public GHZ measurement result AP and peoples know about attempting of Eve.

Detecting Eve's presence requires the comparison of fewer bits. The probability of the result of AP to be $n$ times false is $$\frac{1}{2^n}\left( 1+ 2+\dots+2^{n-1}\right)=\frac{2^n-1}{2^n}. $$
In all these cases peoples observe the brochen results of AP and remove the telecommunication, while Eve cannot discover the secret AS.

\section{Illustration with holographic softwares}
In this section we review the work of A. Jaffe, Z. Liu, and A. Wozniakowski \cite{jaffeetal} involving the softwares of sharing problem. The pictures are taken from their work.

First let us remind that the measurement can also be produced with the code as shown in Figure 4 
\begin{figure}[!ht]
			\centering
				\includegraphics[scale=0.4]{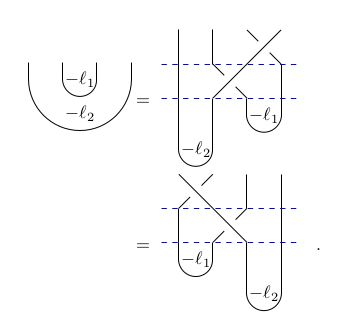}
                             \includegraphics[scale=0.4]{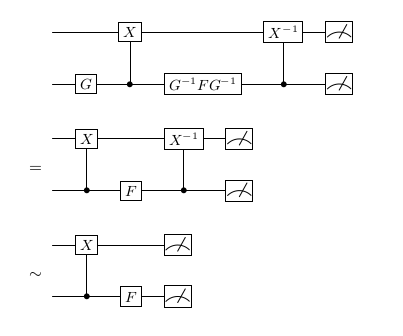}
                     \caption{Measurement and corresponding code}
\end{figure}

This kind of scheme can then be applied to 1-qubit-teleport problem in Figure 5 
\begin{figure}[!ht] 
			\centering
				\includegraphics[scale=0.4]{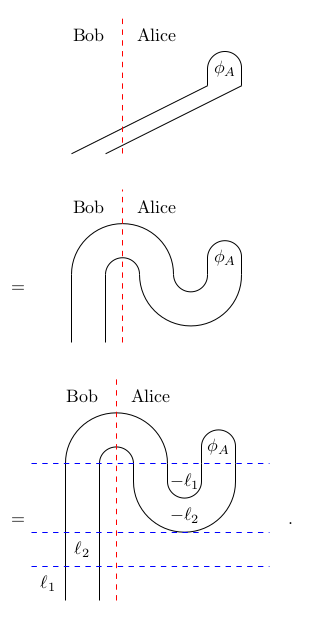}\qquad
                             \includegraphics[scale=0.4]{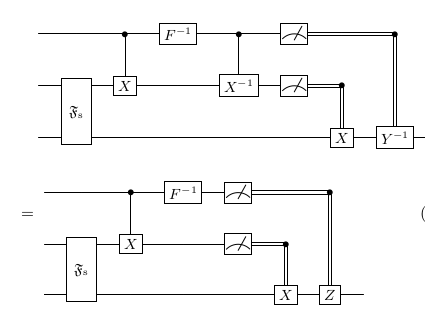}
                     \caption{1-qubit-teleport problem and corresponding code}
\end{figure}

The secret sharing problem between Alice, Bob and Carol can be illustrated as the holographic software and code in Figure 6.
\begin{figure}[!ht]
			\centering
				\includegraphics[scale=0.4]{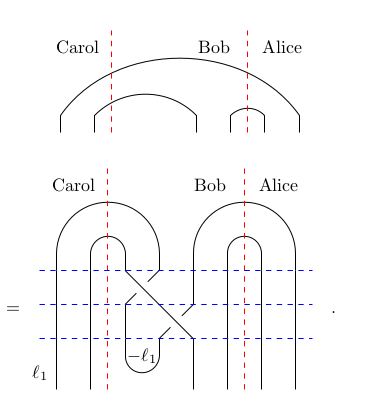}\quad
                             \includegraphics[scale=0.4]{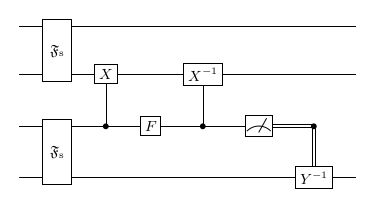}
                     \caption{Secret sharing problem between Alice, Bob and Carol illustrated as the holographic software and corresponding code}
\end{figure}
The scheme for Alice-Bob-Carol sharing is easily generalized as BVK code for n-partitie-sharing problem  as shown in Figure 7
\begin{figure}[!ht] 
			\centering
				\includegraphics[scale=0.4]{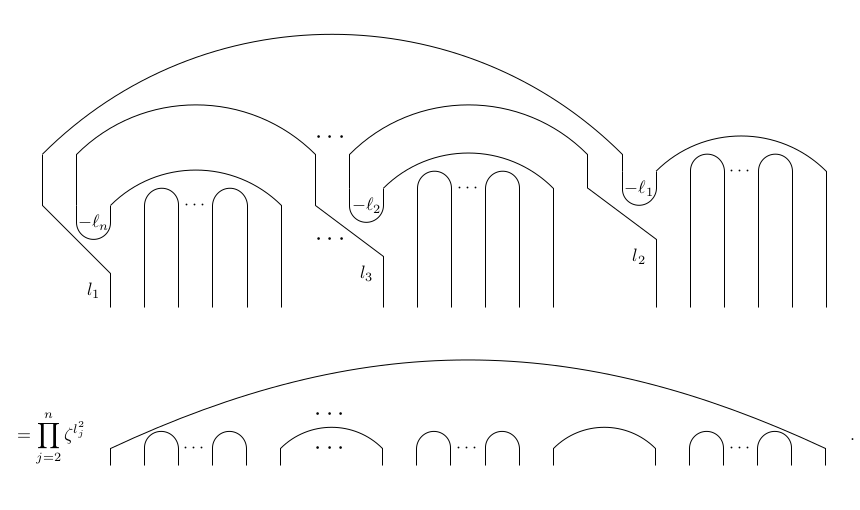}\quad
                             \includegraphics[scale=0.4]{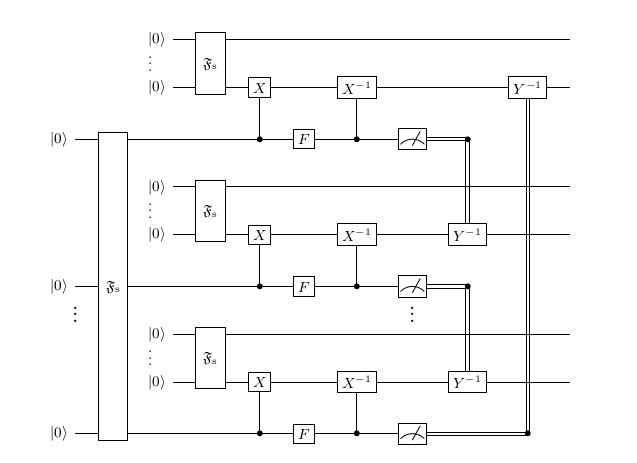}
                     \caption{n-partitie-sharing problem and corresponding BVK code}
\end{figure} 

\section{Conclusion} We proposed to use the Fourier transform measurements for entangled GHZ-state of qubits. The result is independent of indicating the positions and directions, as used in \cite{zengetall}. The result is certainly true also for qudits in place of qubits. The codes can be illustrated by the holographic software and corresponding codes.

\end{document}